\documentclass[preprint,12pt]{elsarticle}

\usepackage{amssymb,amsthm,color,amsmath,graphicx,url}

\newcommand{\rcom}[1]{#1}

\newcommand{\chris}[1]{}

\begin{document}

\begin{frontmatter}
\journal{Pattern Recognition Letters}


\title{A better Beta for the $H$ measure of classification performance}

\author{D.J. Hand, \url{d.j.hand@imperial.ac.uk}, \\
C. Anagnostopoulos, \url{canagnos@imperial.ac.uk}, \\ Department of Mathematics, South Kensington Campus,\\ Imperial College London, London SW7 2AZ}
\maketitle
\begin{abstract}
The area under the ROC curve is widely used as a measure of performance of classification rules.  However, it has recently been shown that the measure is fundamentally incoherent, in the sense that it treats the relative severities of misclassifications differently when different classifiers are used.  To overcome this, \cite{hand2009,hand2010a} proposed the $H$ measure, which allows a given researcher to fix the distribution of relative severities to a classifier-independent setting on a given problem. This note extends the discussion, and proposes a modified standard distribution for the $H$ measure, which better matches the requirements of researchers, in particular those faced with heavily unbalanced datasets, the $Beta(\pi_1+1,\pi_0+1)$ distribution. 
\end{abstract}



\end{frontmatter}

\section{Introduction}
\label{sec:intro}

The aim of supervised classification is to construct a rule which will allow one to assign objects to one of $M$ classes, on the basis of vectors of descriptive features of those objects.  The rule will be constructed using a `training' set (machine learning and pattern recognition terminology) or `design' set (statistics terminology) of data which includes both descriptive vectors and true classes for a sample of objects.  In this note we shall take $M=2$, the most important special case.

Typically, the construction of a classification rule is split into two parts: \rcom{first,} constructing a mapping from the descriptive feature vector, {$\mathbf{x}$,} to a `score' on a univariate continuum, {$s(\mathbf{x}) \in \mathbb{R}$}; \rcom{and second,} choosing a threshold, $t$, with which the score is compared, such that objects are assigned to class $1$ (also known as \emph{positive}) if their score is greater than $t$ and otherwise to class $0$ (also known as \emph{negative}).  In most real problems, however, perfect separation between the classes is not possible: there is no threshold $t$ such that all class $0$ objects score less than or equal to $t$ and all class $1$ objects score greater than $t$.  That is, the distributions of the scores of the two classes, {given by 
\[
F_1(t) = P(s(\mathbf{x}) < t \mid 1), \; F_0(t) = P(s(\mathbf{x}) < t \mid 0)
\]}
will in general have overlapping support.  Among such imperfect classifiers, the question is raised as to which one can claim to be `better' - in a suitable sense.  A great many ways of measuring the relative performance of classification rules have been proposed. These include measures based on misclassification counts, such as \emph{sensitivity}, \emph{specificity}, \emph{positive} and \emph{negative predictive value}, \emph{proportion correct} and its complement \emph{error rate}, the \emph{$F$ measure}, and others (for discussion, see, for example, \cite{hand1997,hand2001,hand2005,hand2011,zhou2002,pepe2003,gonen2007,krzanowski2009}).  {Measuring performance on the basis of misclassification counts is complicated by the fact that, for each classifier, the threshold $t$ remains a free parameter that crucially affects performance, as it negotiates the tradeoff between \emph{false positive} and \emph{false negative} classifications: a very large value for $t$ will tend to classify almost all objects as class $0$, whereas a very low value for $t$ will tend to err on the other side, classifying most objects as class $1$. This is often illustrated by the Receiver Operating Characteristic (ROC) Curve, which is obtained by plotting the True Positive Rate (TPR) of the classifier, given by $1-F_1(t)$,  against its False Positive Rate (FPR), given by $1-F_0(t)$,  for all possible values of the threshold. ROC curves extend from $(0,0)$ to $(1,1)$  by gradually sacrificing false negatives for false positives.}

{The majority of the performance measures mentioned above handle the dependence of performance on the free parameter $t$ either by requiring that the threshold should be fixed by the user, or by implicitly specifying a value for it.} {For instance, the error rate of a classifier is given by setting the {threshold} to the value that minimises the total number of erroneous classifications:
\[
ER = \underset{t}{\text{min}}\, \{\pi_0(1-F_0(t)) + \pi_1 F_1(t)\}
\]
where $\pi_0$ and $\pi_1$ are the respective proportions of class $0$ and class $1$ objects in the population. Justifying this, or any other predetermined choice of threshold is difficult, mainly because the relative importance of the two different types of misclassification errors will in general depend on the problem, so that the threshold may often not be chosen until the rule is applied in practice.} To sidestep this problem, the Area Under the ROC curve (\textit{AUC})  measure (also called the $c$-statistic, and equivalent to the \emph{Gini coefficient}, which is a chance-standardised version) is very widely used. {The \textit{AUC} can be intuitively motivated by the observation that if one ROC curve lies strictly above another, then the respective classifier performs better at all threshold levels. This suggests the area under the curve as a possible scalar summary of \emph{aggregate} performance.} The \textit{AUC} however admits several other interpretations.  It is the probability that a randomly chosen member of class $0$ will yield a score lower than a randomly chosen member of class $1$ -- and from this it follows that it is the same as the test statistic used in the Mann-Whitney-Wilcoxon two sample nonparametric test to compare two distributions.  It is the average sensitivity if specificity values are chosen uniformly, and the average specificity if sensitivity values are chosen uniformly.  It is  also a linear transformation of the proportion correctly classified if the threshold is randomly chosen from an arbitrary linear combination of the class score distributions \citep{hand2011b}, with the coefficients of the transformation being functions of the class priors (the relative proportions of objects belonging to each class, denoted $\pi_0$ and $\pi_1$ for classes $0$ and $1$ respectively in what follows).  \rcom{This interpretation is particularly revealing as }it shows explicitly that the \textit{AUC} is an aggregate or portmanteau measure, equivalent to integrating over a range of possible values for the threshold $t$. 

Unfortunately, in a series of papers, \citep{hand2009, hand2010a, hand2011b} it was demonstrated that when the classification of an object is to depend only on the score of that object and the threshold with which it is to be compared (and not, for example, on the scores of other objects) then the area under the ROC curve is an incoherent performance measure, in the sense described immediately below.

For a given threshold $t$, the four probabilities in the cross-classification table of true class by predicted class are constrained by two relationships: that the total proportion in class $0$ is $\pi_0$ and the total proportion in class $1$ is $\pi_1 = 1-\pi_0$. This thus leaves two degrees of freedom, which have to be reduced to one to provide a univariate measure which can be used to compare classifiers.  Different performance measures effect this reduction in different ways. For instance, the error rate is simply the weighted average, with weights given by the class proportions in the population, of the proportions of each class misclassified; the \textit{KS} statistic is (proportional to) the minimum (by choice of $t$) of the overall proportion misclassified if the proportions misclassified in each class are equally weighted; etc.. The \textit{AUC} sidesteps the requirement to specify $t$ by integrating a weighted misclassification rate over a distribution of $t$ values, as described above.  However, \cite{hand2009,hand2010a} showed that, when considered in terms of the ratio of the severity of misclassifying a class 0 objects as class 1 to the severity of misclassifying a class 1 object to class 0, this implies that different classifiers adopt different distributions for this ratio. 
This is nonsensical, since this ratio is a property of the problem, not the instrument used to make the classification: the distribution of the ratio of misclassification severities must be the same for all classifiers applied to a given problem.  \cite{hand2011b} reformulated the argument in terms of calibrated score distributions, which allowed them to avoid the need to introduce reference to misclassification costs.

\section{Choosing the threshold distribution}

To overcome the deficiency of the \textit{AUC} described above, \cite{hand2009,hand2010a} defined an alternative measure, the $H$ measure, which proposed using a fixed relative misclassification severity distribution. {We provide here a very brief outline of the measure, and refer to \cite{hand2009} for more details.} Let $c$ in $[0,1]$ denote the `cost' of misclassifying a class $0$ object as class $1$, and $1-c$ the cost of misclassifying a class $1$ object as class $0$. {Consider then the following \emph{loss} function, which represents the total cost incurred:
\[
L(c;t) = c\pi_0 (1-F_0(t)) + (1-c)\pi_1 F_1(t)
\]
In this context it is natural to choose the threshold $t$ to minimise the total loss, yielding a \emph{minimum loss} of $L(c;T_c)$, where
\[
T_c = \underset{t}{\text{argmin}}\,L(c;t)
\]
In this setup, the threshold is no longer a free parameter, but rather fully determined by the normalised cost, $c$. However, as we explained earlier, fixing the cost $c$ to a single value in advance is too strict a requirement. Instead, it is more realistic to specify a \emph{distribution}, $w(c)$, over different values of $c$:
\[
L = \int_c L(c;T_c)w(c)dc
\]}
And this is exactly what the \textit{AUC} does. However, the \textit{AUC} requires that $w(c)$ differs between different classifiers, so that different measures are being used to evaluate different classifiers. In contrast, the $H$ measure requires that the same w distribution is used for all classifiers. 

Although for the $H$ measure the distribution $w(c)$ is fixed -- in the sense that any given researcher should choose a distribution and use that for all classifiers being applied on the given problem -- it is not appropriate to objectively specify a universal distribution that all researchers should use for all problems. This is because different researchers may well have different beliefs about the relative misclassification severities, and because it is entirely likely that different problems will merit different distributions.  There thus remains an intrinsic and fundamental arbitrariness about the choice of $w(c)$.

To tackle this, \cite{hand2009} suggested that the value of the $H$ measure should be reported for two distinct relative severity distributions.  One should be a subjective distribution, chosen by each researcher for each problem (but the same for all classifiers applied by that researcher to that problem, of course).  The other should be a universal standard, and \cite{hand2009} proposed the $Beta(2,2)$ distribution.  However, in response to experience from a number of researchers in using the $H$ measure on a wide variety of problems we would now like to propose a modified universal standard.

In many problems the class sizes are extremely unbalanced.  For example, one of the researchers who contacted us had $\pi_1 = 0.024$ and another had $\pi_1 = 0.00032$.  In such cases, it would be rare that one would want to use a symmetric relative severity distribution because of the symmetry this implies about the way the classes are treated. Instead, one would probably want to treat misclassifications of the smaller class as more serious than those of the larger class: \rcom{if they are treated as of equal severity then very little loss would be made by assigning everything to the larger class.}  To take an example, in credit card transaction fraud detection \citep{hand2008b}, most transactions are legitimate -- the class sizes are very unbalanced.  Moreover, misclassifying a legitimate transaction as fraudulent may incur only the cost of an investigatory phone call, plus some small fraction of the associated wage bill of the employee making the call, as well as a small part of the bank's infrastructure costs.  But all these are likely to add up to far less than the cost of misclassifying a fraudulent transaction as legitimate -- which could easily run into the thousands of dollars.

Recognising that one would not want to use a symmetric distribution, and not wishing to choose one subjectively (despite our recommendation that they should do so, noted above), the researchers sought another standard alternative.  In response to this, we propose the following.

Consider first the \textit{KS} statistic. This chooses $c$ so that the cost incurred if all the class $0$ objects and none of the class $1$ objects are misclassified, is equal to the cost incurred if all the class $1$ objects and none of the class $0$ objects are misclassified.  This results in a larger misclassification cost for each of the objects from the smaller class, and equal costs if the class sizes are equal.  In particular, of course, in cases when the classes are very unbalanced it gives dramatically larger costs to misclassifications from the smaller class.  In the fraud detection example above, if there are $1000$ legitimate transactions to every fraudulent credit card transaction (which is in fact the order of magnitude of the ratio in such problems), then the cost attributed to misclassifications of a fraudulent transaction is set at $1000$ times the cost of misclassifying a legitimate transaction.  In general, the \textit{KS} achieves this effect by setting $c=\pi_1$ and $1-c = \pi_0$.  It is the essence of the $H$ measure (and indeed the principle underlying the \textit{AUC}) that we want to avoid choosing a single fixed value of $c$, and instead pick a distribution.  We therefore propose choosing a distribution such that the mode of the relative misclassification severity distribution in the $H$ measure should be at $c=\pi_1$.  This means that, for example, in highly unbalanced situations, one regards it as more likely that misclassifications from the smaller class will be more serious than misclassifications from the larger class. 

\begin{figure}
\includegraphics[width=\textwidth]{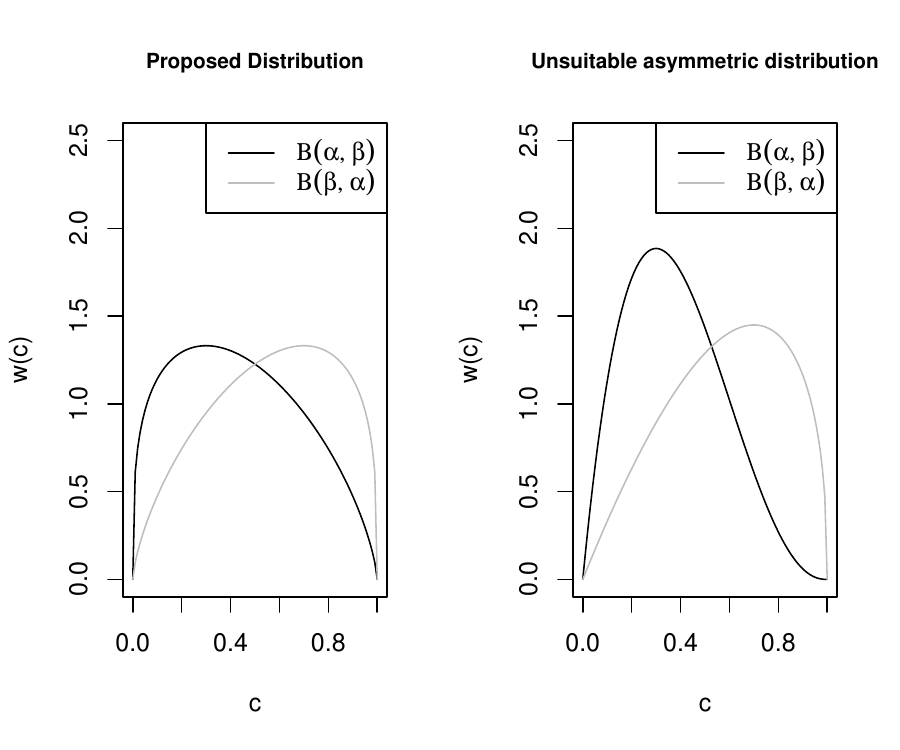}
\caption{\label{fig:symmetry}The proposed distribution's dependence on $(\pi_0,\pi_1)$ must be such that employing the pair $(\pi_1,\pi_0)$ instead yields a reflected version of the distribution.}
\end{figure}

\begin{figure}
\includegraphics[width=\textwidth]{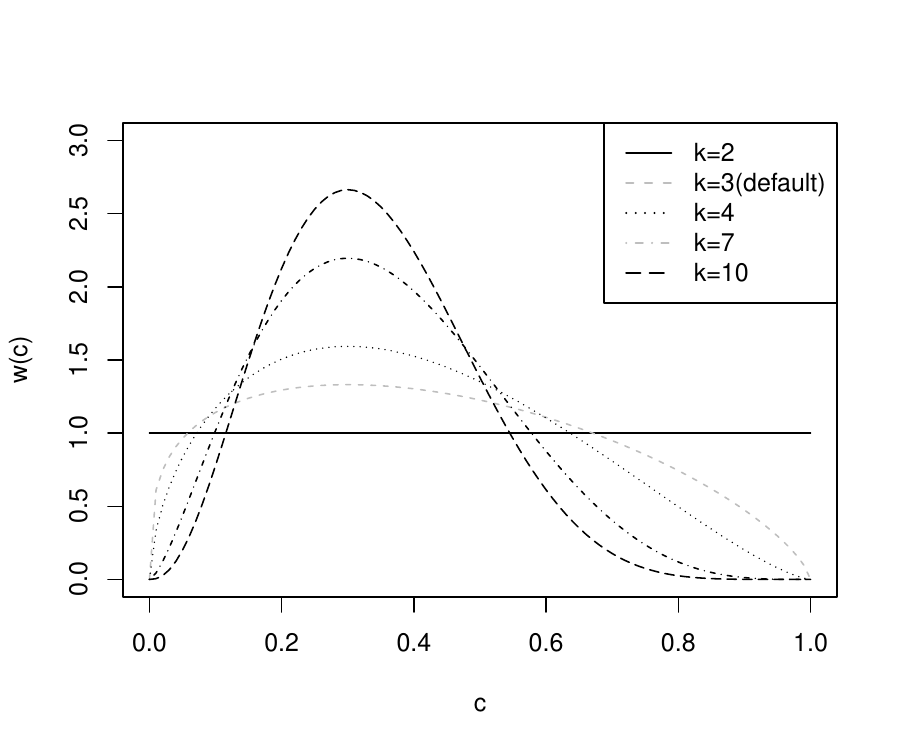}
\caption{\label{fig:width}The parameter $k$ controls the dispersion about the mode, so that larger values of $k$ may be used to reflect higher degrees of certainty about the setting $c = \pi_1$.}
\end{figure}

{For a $Beta(\alpha,\beta)$ distribution with $\alpha > 1$ and $\beta > 1$, the mode is: 
\[
\frac{\alpha-1}{\alpha+\beta-2}
\]
We can set this mode equal to $\pi_1$ in several ways. For instance, we may set
\[
\beta = 1 + (\alpha-1)\frac{\pi_0}{\pi_1}.
\]
leaving open the choice of $\alpha$, with $\alpha=2$ being a reasonable default value on the grounds that it gives a unimodal distribution which is not too extreme.  The result is a $Beta(2,\pi_1^{-1})$ distribution, which, for fully balanced problems with $\pi_0 = \pi_1$,  reduces to $Beta(2,2)$. Nevertheless, 
this distribution suffers from the disadvantage that it treats its two parameters $\alpha$ and $\beta$ asymmetrically. To understand why this is undesirable, consider $Beta(\alpha(\pi_0,\pi_1), \beta(\pi_0,\pi_1))$ to be the general form of a Beta distribution whose parameters are selected using the class priors. Since switching the class labels around would have the effect of replacing $(\pi_0,\pi_1)$ with $(\pi_1,\pi_0)$, and $c$ with $1-c$, we must require of our cost distribution that:
\begin{equation}
\label{eq:symmetry_cons}
c \sim Beta(\alpha(\pi_0,\pi_1), \beta(\pi_0,\pi_1)) \; \Rightarrow \; 1-c\sim Beta(\alpha(\pi_1,\pi_0), \beta(\pi_1,\pi_0))
\end{equation}
Noting that, for all Beta distributions,
\[
c \sim B(\alpha,\beta) \Rightarrow 1-c\sim Beta(\beta,\alpha)
\]
we immediately observe that property (\ref{eq:symmetry_cons}) does not hold of $Beta(2,\pi_1^{-1})$. Instead, we may enforce symmetry by setting $\alpha+\beta = k$. To place the mode to a value $\tilde{c}$, we then need to specify $\alpha$ and $\beta$ as follows:
\begin{equation}
\label{eq:better}
\alpha = (k-2)\tilde{c} + 1, \; \beta = (k-2)(1-\tilde{c}) + 1, \text{ for }k \geq 3
\end{equation}
Different values of $k$ in (\ref{eq:better}) make the proposed distribution narrower or wider, as illustrated in Figure \ref{fig:width}, but leave the mode unaffected. A sensible default value for $k$ is $3$, which, together with $\tilde{c} = \pi_1$, yields the $Beta(\pi_1+1,\pi_0+1)$ distribution as the default universal standard for the $H$-measure. The symmetry requirement is illustrated in Figure \ref{fig:symmetry}, where a $Beta(2,\pi_1^{-1})$ and the proposed default distribution are plotted alongside the respective distributions obtained by switching the shape parameters around.}


Clearly, in contexts where domain knowledge suggests a reasonable \emph{approximate} `guess' $\tilde{c}$ for the normalised cost $c$, this value may be used in equation (\ref{eq:better}) in place of $\tilde{c} = \pi_1$ to place the mode on the cost estimate. In such a case,  the parameter $k$ controls the degree of certainty about that estimate (see Figure \ref{fig:width}). In certain contexts it may be easier to elicit an expert opinion about the \emph{relative severity ratio} $r$ instead, i.e., the ratio of the costs of the two types of misclassification errors given by $r = \frac{c}{1-c}$. The quantity $r$ measures how much more severe misclassifying a class 0 instance is than misclassifying a class 1 instance. Given a `guess' $\tilde{r}$ and inverting its relationship with $c$, one obtains $\tilde{c} = \frac{\tilde{r}}{1+\tilde{r}}$ which may be employed in (\ref{eq:better}) as before, to produce a distribution whose single mode is placed on the expert guess for the relative misclassification cost. In either case, the proposed construction reduces the burden to the individual researcher of fully specifying a $Beta(\alpha,\beta)$, and will hopefully encourage users to deploy domain knowledge whenever possible (which we argue is possible more often than not), making full use of the expressive power of the $H$ measure. However, in the absence of such domain knowledge, and for the purpose of making available a universal standard, we propose here that the setting $c = \pi_1$ (i.e., $r = \pi_1/\pi_0)$ is in fact a better default than our earlier suggestion $c = 0.5$ (i.e., $r=1$) which underlied the $Beta(2,2)$.

\section{Conclusion}

\cite{hand2009, hand2010a} showed that, when classifications were to be based solely on the score of an object and the threshold with which it was to be compared, the \textit{AUC} was fundamentally incoherent in the sense that it treated different classification rules differently: it is equivalent to letting the choice of measuring instrument depend on the object being measured.  To overcome this problem, he proposed the $H$ measure, which fixes the distribution in a classifier-independent manner, so leading to an invariant measure.  This distribution cannot be chosen in a fully objective way across all problem domains, as it will depend on the problem and the researcher's beliefs about the consequences of the different kinds of misclassification, but the $H$ measure fixes it for a given researcher working on a given problem.  For this reason, \cite{hand2009, hand2010a} proposed that the $H$ measure with two forms of distribution should be reported for each study: \rcom{first,} a subjective distribution based on the researcher's beliefs; \rcom{second,} a universal standard distribution.  For the latter, he suggested a $Beta(2,2)$ distribution.

Now that experience with the $H$ measure is accumulating, and based on correspondence with researchers throughout the world who have used it, \rcom{it seems more suitable to introduce a standard distribution with an asymmetric relative cost distribution for unbalanced problems, that also reduces to the $Beta(2,2)$ distribution for balanced problems.} This paper introduces exactly such a candidate, the $Beta(\pi_1+1,\pi_0+1)$ distribution.

\bibliographystyle{plain}







\end{document}